# Chemical control of polymorphism and ferroelectricity in PbTiO$_3$ and SrTiO$_3$ monolayers and bilayers

Shaowen Xu,[a,b] Jeffrey R. Reimers,[a,c†] Fanhao Jia[a,d] and Wei Ren[a†]

Layers of perovskites, found in 3D materials, 2D heterostructures, and nanotubes, often distort from high symmetry to facilitate dipole polarisation that is exploitable in many applications. Using density-functional theory calculations, ferroelectricity in bilayers of the 2D materials PbTiO$_3$ and SrTiO$_3$ is shown to be controlled by bond breakage and formation processes that act as binary switches. These stacking-dependent processes turn on and off as a function of relaxation from high-symmetry structures and the application of biaxial strain, and their concerted rearrangements lead to low energy barriers for ferroelectric polarisation switching. Structures with symmetry intermediate between high-symmetry octahedral forms and low-symmetry ferroelectric forms are identified, allowing the intrinsic processes associated with traditional "ferrodistortive" and "antiferrodistortive" distortions of TiO$_6$ octahedra to be identified. Ferrodistortive-mode activity is shown to be generated by the simultaneous application of two different types of curvilinear antiferrodistortive motions. In this way, four angular variabes control polarisation switching through the concerted making and breaking of chemical bonds. These subltities make the polarisation sensitive to chemical-environment and temperature effects that manipulate strain and structure, features exploitable in futuristic devices.

## 1. Introduction

Artificially layered atomistic superlattices constructed by stacking ferroelectric and paraelectric oxides provide possibilities for designing multifunctional materials.[1-11] These heterostructures usually possess enhanced properties compared to a simple combination of the parent compounds, which can be explained by the effects of lattice mismatch and dimensionality, as well as competition between different phases and coupling of structural instabilities.[12] An important current challenge is the development of simple model systems that can reveal the basic interactions controlling the properties of these complex materials.

Perovskites, of form ABO$_3$, have intrinsic octahedral coordination about atom B. Of particular interest, this symmetry can be lowed, allowing for ferroelectric polarisation. Two primary types of distortions have been identified:[7,13]

- Ferroelectric distortions (FE) that are typically characterised in terms of translations of the A atom amidst the surrounding BO$_3$ lattice, or else translations of the B atom with respect to its oxygen octahedron.[1,14] These can induce net in-plane and/or out-of-plane electric polarisation[15] that is called "proper" ferroelectricity.[16-18]
- Antiferrodistortive (AFD) effects involving apparent rotations of BO$_6$ octahedra amidst the surrounding A lattice.[13,17,19-21] Such rotations of a nominally symmetric species are not expected to induce polarisation as they involve opposing motions of symmetrically related atoms. This notation parallels the use of "antiferromagnetism" to indicate how symmetrically equivalent atoms in some smaller primitive cell develop opposing properties in the expanded unit cell of the material in question.[22-24]

Ferroelectric and antiferrodistortive effects were originally perceived as competing against each other,[25,26] but now situations are known in which FE and AFD effects are regarded as co-operating to enhance ferroelectricity.[7,27] Also, coupling of AFD motions to other motions, and changes in atomic charges, can result in intrinsic generation of ferroelectricity, known as "improper" ferroelectricity.[16] In this work, for the 2D perovskite materials in question, ferroelectric distortions are shown to arise as a *consequence* of the operation of two *simultaneous* AFD effects, which indeed is a type of cooperativity.

In general, the apparent interplay of the FE and AFD effects can have profound consequences in materials, including 3D solids of pure perovskites, mixed 3D samples, sophisticated heterostructures, 2D materials, and 1D nanowires. Classic materials used to consider this include those made of PbTiO$_3$ (PTO) or SrTiO$_3$ (STO) or combinations thereof.[17] In samples prepared that show short-range crystalline order, the ground-state has been described in terms of "an interfacial induced form of improper FE" that manifests a very large and temperature-independent dielectric constant.[17] This milestone observation has inspired a series of studies of interfacial effects,[28-32] particularly the coupling of mechanical/electric boundary conditions and ferroelectric properties that have been rationalised in terms of, e.g., FE-AFD-strain coupling.[7,33]

In superlattices prepared showing only long-range crystalline order (tens of unit cells), various polar domain patterns have been observed. For example, ferroelectric polar vortices and antivortex have been experimentally realised in superlattices made from PTO and STO layers, which can be controlled by external strains, thermal and electrical stimuli.[34-37] Ferroelectric flux-closure quadrants were found in PTO/STO multilayers that were obtained by tuning the thickness ratio of adjacent PTO layers.[38-40] Also, PTO/STO superlattices have been shown to display in-plane polarisation at room temperature.[41] PTO itself also shows related interesting effects including polymorphism,[42,43] and can be formed into small ferroelectric nanotubes with many possible applications.[44-48]

An avenue for understanding the properties of bulk materials and large heterostructures is through the consideration of the basic structural units, comprising of single ABO$_3$ layers, and how they combine to form bilayers. Indeed, component 2D freestanding perovskite oxides, down to the monolayer limit, such as STO and BiFeO$_3$, have been synthesised using a water-soluble sacrificial buffer layer.[9] Super-elastic ferroelectricity with ~180° recoverable folding and ~10% strain during the in situ bending was found in single-crystal BaTiO$_3$ membranes.[49] These 2D perovskites were reported to present unusual ferroelectricity that was induced by surface effects[50] or by spin-lattice coupling.[39] Further, 2D mixed oxide perovskite heterostructures have been fabricated by a remote epitaxy on the top of graphene interlayers, including Pb(Mg$_{1/3}$Nb$_{2/3}$)O$_3$-PTO/ SrRuO$_3$ and CoFe$_2$O$_4$/Y$_3$Fe$_5$O$_{12}$ heterosturctures.[51] An electronically reconfigurable 2D electron gas was then preserved in freestanding LaAlO$_3$/STO membranes.[52] Large magnetoelectric coupling was found in the multiferroic oxide La$_{0.7}$Sr$_{0.3}$MnO$_3$/SrRuO$_3$ that assembled via epitaxial lift-off.[53]

In other fields, organic layers have also been used to nucleate 2D perovskite thin-film transistors,[54] indicating the importance of surface chemistry. The net polarisation can be controlled by surface effects.[55] They can be critical in controlling the properties of perovskite solar cells,[56] and can also be critical to stabilizing perovskite light-emitting diodes.[57] In applications such as catalysis, surface effects associated with unterminated chemical bonding are also known as one of the fundamental determinants of reactivity.[58] Considering freestanding monolayers and bilayers provides means for revealing the factors controlling surface effects and their significance for general heterostructures.

We consider properties of PTO and STO monolayers and their bilayers. Computational means are applied using the PBESol density functional[59] to optimise and vibrationally characterise structures, as well as the asymptotically corrected CAM-B3LYP functional[60-64] applied to increase the accuracy of the perceived energetics along with the D3(BJ) dispersion correction.[65] PTO and STO monolayers, their single-component bilayers PTO/PTO and STO/STO, and their mixed bilayers STO/PTO and PTO/STO are studied.

Feasible structures for the monolayers and bilayers could take the form AO-BO$_2$ and AO-BO$_2$-AO-BO$_2$, respectively, in which AO and BO$_2$ present as atomic planes within each layer, see Fig. 1. Two other extended forms are also possible: AO-BO$_2$-AO monolayers with AO-BO$_2$-AO-BO$_2$ biyalers, or else BO$_2$-AO-BO$_2$ monolayers with BO$_2$-AO-BO$_2$-AO-BO$_2$ biyalers. In all cases, the top and bottom atomic planes will present undercoordinated metal atoms, either Pb or Sr if the outside plane is AO and Ti if the outer plane is BO$_2$. We initially consider all possibilities for the monolayers, concluding that the extended structures are less likely, and then proceed using only AO-BO$_2$ monolayers and AO-BO$_2$-AO-BO$_2$ bilayers (Fig. 1). Extended structures are also considered unlikely based on: (i) observed polarisation data for STO and BaTiO$_3$ films,[9] (ii) they may not be consistent with observed stoichiometric properties of analogous Ni films,[66, 67] (iii) they are not preferred in analogous Ta films,[68-70] and (iv) they have predicted metallic properties[68, 71] that have not observed in related experiments. Prior DFT simulations have also predicted the mechanical stability of STO, BaTiO$_3$ and other AO-BO$_2$ monolayers.[10, 72, 73]

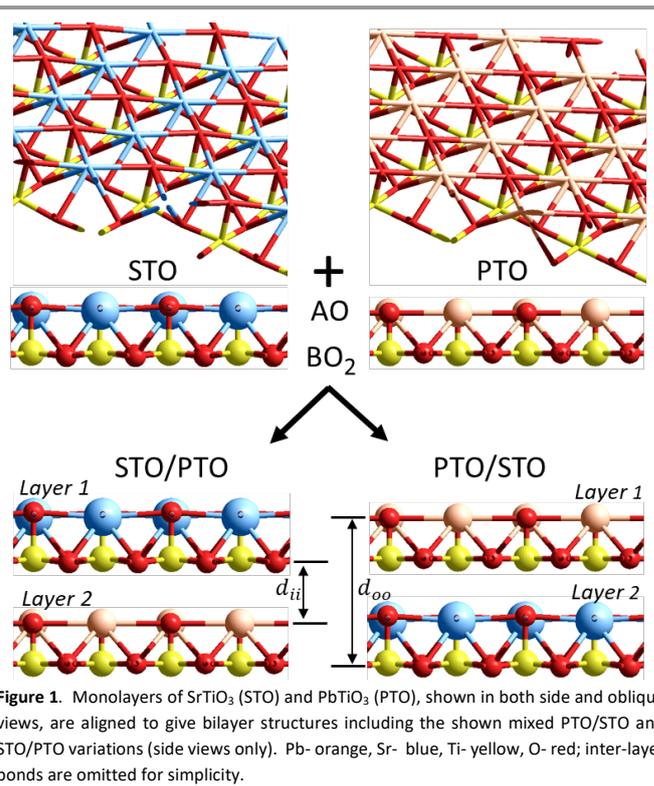

**Figure 1**. Monolayers of SrTiO$_3$ (STO) and PbTiO$_3$ (PTO), shown in both side and oblique views, are aligned to give bilayer structures including the shown mixed PTO/STO and STO/PTO variations (side views only). Pb- orange, Sr- blue, Ti- yellow, O- red; inter-layer bonds are omitted for simplicity.

This work then focuses on the making and breaking of the chemical bonds as different types of distortions occur in the AO-BO$_2$ and AO-BO$_2$-AO-BO$_2$ monolayers and bilayers. As a result, different distortion types are seen to arise as a consequence of chemical bond breakage and formation, with, in most cases, bonds acting as binary switches inside the materials. This differs significantly from traditional interpretations of the role of chemical effects in determining ferroelectricity, which instead focus only on whether bonds are primarily ionic or covalent in nature.[1, 7] To reveal the changes in the bonding topology that drive ferroelectricity, we optimise structures that are stationary points on the ground-state potential-energy surface with ever-decreasing symmetry. This reveals a cascade of chemical reactions that leads from high-symmetry atomic arrangements to the final low-energy ferroelectric polymorphs. Concerted rearrangements of the bonds then lead to low energy pathways for ferroelectric switching. Also, series of biaxial strains are applied to all monolayer and bilayer structures, revealing how the critical chemical-bonding binary switches operate.

## 2. Methods

### 2.1 Structural and symmetry representation.

As none of the structure considered embody symmetry features that are orthogonal to the 2D plane, all symmetry information is described simply in terms of the 17 plane groups.[74] As a result, structural properties are interpreted in terms of reflection planes and guide-reflection planes, as well as two-fold and four-fold rotation axes, that are all normal to the 2D plane.

Structurally, the monolayers and bilayers are modelled using mostly $(\sqrt{2} \times \sqrt{2})R45°$ supercells of monolayers extracted from the $Pm\bar{3}m$ bulk structure of SrTiO$_3$.[75] This transformation produces a 45° in-plane rotation of the original *a* and *b* crystallographic axes to produce new axes *x* and *y*; the *c* axis is unperturbed and becomes the new *z* axis. For PTO/STO bilayers, for example, each of these unit cells has the composition $Pb_2Sr_2Ti_4O_{12}$, containing 20 atoms. In addition, calculations for the PTO/PTO **OCT** and **Q** structures (see later), as well as for PTO/STO **OCT**, **P**, **Q**, and $\mathbf{2S^\theta}$, are performed on 40-atom $(2 \times 2)$ supercells to check for stability and convergence of free-energy calculations, as well as to better expose symmetry reflection planes in *cmm* and *cm* symmetry. Also, PTO/STO **Q** is considered using a large 80-atom $(2\sqrt{2} \times 2\sqrt{2})R45°$ supercell to check for medium-range ordering. The bilayers are modelled in 3D-periodic cells of height *z* = 30 Å, introducing a large vacuum region of ca. 24 Å between bilayer images. Polarisations calculated in the *z* direction conserve the centre of mass, whereas in-plane polarisations are origin independent.

### 2.2 Energy calculations, electronic analyses, and geometry optimisation.

Calculations based on density-functional theory (DFT) are performed using the Vienna *ab initio* Simulation Package (VASP),[76, 77] with the electron-nuclear interactions described using projector augmented wave (PAW) pseudopotentials.[78] Most calculations are performed using the generalised gradient approximation (GGA) in the form proposed by Perdew, Burke, and Ernzerhof (PBE)[79] after revision to enhance calculations of the properties of solids (PBEsol).[59] The convergence criterion on the energy is set to at most $10^{-6}$ eV, and atomic positions were fully relaxed until the maximum force on each atom was less than $10^{-3}$ eV/Å. The precision of the total energy was set to be $10^{-8}$ eV. Some single-point energy calculations are made using the asymptotically corrected CAM-B3LYP hybrid density functional,[60-64] corrected for dispersion using the D3(BJ) approximation.[65]

In most calculations, the VASP "PBE" pseudopotentials were used that explicitly embody the 4*s*, 4*p*, 5*s* electrons for Sr, the 5*d*, 6*s*, and 6*p* electrons for Pb, the 3*s*, 3*p*, 3*d*, 4*s* electrons for Ti, and the 2*s*, 2*p* electrons for O. Alternatively, for the CAM-B3LYP calculations and the sensitive calculations reported in Fig. 9, the more accurate "GW" pseudopotentials were used instead. Changing the pseudopotential resulted in structural energy differences of at most 3 meV.

For calculations using $(\sqrt{2} \times \sqrt{2})R45°$ supercells, a 6×6×1 Γ-centred Monkhurst-Pack *k*-mesh was used. This is reduced to 4×4×1 for $(2 \times 2)$ supercells and 3×3×1 for $(2\sqrt{2} \times 2\sqrt{2})R45°$. The energy cutoff for the plane-wave basis set is chosen to be 600 eV. Such a large value is required as the energy changes of interest are often small. For the six PTO/STO structures of greatest interest, in test calculations, relative energy differences were re-evaluated using an energy cutoff of 900 eV and *k*-meshes twice the size in each direction, repeated for both $(\sqrt{2} \times \sqrt{2})R45°$ and $(2 \times 2)$ supercells. Based on these results, the numerical error in calculated energy differences is determined to be 1 meV.

Structures were optimised within specified plane groups by enforcing symmetry operators using the VASP ISYM command. Small deviations from the desired symmetry were sometimes obtained and then removed. All optimised structures, as well as basic information concerning the calculation parameters and results, are provided in Electronic Supporting Information (ESI) part S3.

### 2.3 Vibrational analyses, structure characterisation, and free energies.

PBESol phonon analyses to obtain vibration frequencies and normal modes at the gamma point of the vibration Brillouin zone are evaluated using VASP by numerical differentiation of calculated atomic forces. A key purpose for these analyses is the determination of the number of imaginary frequencies so as to characterise optimised structures. As the numerical error associated with the differentiation is sufficiently large that the expected zero-frequency cell-translation modes appeared with frequency magnitudes larger than those of interest, the zero-frequency modes are projected out using the DUSHIN package.[80] This transforms the Hessian matrix into curvilinear internal coordinates, analytically removing the translational modes, and then transforming back into Cartesian coordinates.[81] The resulting vibrational frequencies are then used to calculate free energy differences $\Delta G$ at 298 K from electronic-energy differences $\Delta E$ in the standard way.[82]

Mostly, the minimum-sized appropriate unit cell is used, as well as test calculations using cells of twice the size for PTO/STO materials. Changes of at most 2% are found for $\Delta G^0$ and at most 10% for $\Delta G^{298}$.

### 2.4 Representation of isomerism and isomerisation.

In graphical representations of heterostructures, bonds are shown whenever atomic distances to Pb or Sr are less than 2.8 Å or else atomic distances to Ti are less than 2.2 Å. As shown later in Fig. 5, these definitions result in an unambiguous divide. For all structures calculated, full lists of calculated bond lengths, as well as some critical angles, and provided in ESI Part S1.

### 2.5 Molecular Dynamics simulations of chemical stability.

As tests for chemical stability, PTO/STO and STO/PTO bilayers are simulated at 300 K and 1 bar pressure using molecular dynamics (MD). These simulations ran for 1 ps with a time step of 4 fs, testing to see if significant structural changes could be induced under these conditions.

### 2.6 Calculations of charge polarisation changes.

Internal cell charge polarisations changes between structure pairs are computed by evaluating chemically realistic Bader atomic charges[83] for each structure and considering the atomic displacement vector between the charges. This is done by averaging over symmetrically equivalent structures obtained by applying the glide reflections $(x + 1/2, y - 1/2, z)$ and $(x - 1/2, y + 1/2, z)$ from amongst the *p4m* symmetry-operator set of OCT structures. This ensures that the calculated polarisations embody no quanta of polarisation and are invariant to all translational, rotational, and reflection symmetry operations. The result is that polarisations can be understood in intuitive ways (such as "proper" and "improper" ferroelectricity), avoiding the use of unintuitive Berry-phase-type schemes[84, 85] that also address the critical quanta-of-polarisation and invariance issues. By this procedure, only changes in polarisation from one structure to another can be calculated, and full details of all analyses are given in Excel files in ESI. Therein, the total polarisation change is partitioned into a components associated with changes to the atomic charges and with atomic displacements. As this partitioning is method dependent, it is indicative only.

The polarisation changes are described as an effective induced dipole per unit volume, $P$. This is standard practice for 3D materials, allowing ready adaption of the results into heterostructures and bulk studies. To do this, a standard thickness for each monolayer or bilayer is obtained by summing the layer thickness pertinent to the 3D calculated structures of: PTO and STO of 3.969 Å per layer and 3.945 Å per layer, respectively. Results are presented in µC/cm$^2$, taking 1 e/Å$^2$ = 1602 µC/cm$^2$.

## 3. Results and discussion

### 3.1 Relative stability of AO-BO$_2$, AO-BO$_2$-AO, and BO$_2$-AO-BO$_2$ structural units

Given the interest in the general understanding of how many atomic planes are needed to make stable monolayers carved from ABO$_3$ materials, we consider first the bandgaps predicted for high-symmetry polymorphs, named **OCT**, of PbO-TiO$_2$ (PTO), SrO-TiO$_2$ (STO), and their three-plane counterparts PbO-TiO$_2$-PbO, TiO$_2$-PbO-TiO$_2$, SrO-TiO$_2$-SrO, and TiO$_2$-SrO-TiO$_2$, with the results given in Table 1. The bandgap is a generic indicator of the stability of semiconducting materials to chemical attack.[86, 87] PBEsol[59] is known to systematically understate bandgaps, and so predictions that materials are semiconductors can be taken as being reliable. For STO and PTO monolayers,

PTO is predicted to have a significant bandgap of 2.06 eV, with PbO-TiO$_2$-PbO and TiO$_2$-PbO-TiO$_2$ having much lower bandgaps of 1.43 eV and 1.23 eV, respectively (Table 1). Structures with lower bandgaps are more prone to structural instabilities and their associated chemical reactions.[86, 87] Hence the enhanced bandgap suggests that PTO is the most likely candidate for a stable monolayer structure. STO is predicted to have a significant bandgap of 1.78 eV, but now protecting the exposed Ti face leads to a slightly increased bandgap of 1.97 eV for SrO-TiO$_2$-SrO, with again having only Ti exposed to the outsides in TiO$_2$-SrO-TiO$_2$ leads to a low bandgap of 1.15 eV. It seems reasonable therefore to adopt PTO and STO as the basic monolayer building blocks for heterostructures. The calculated bandgaps for these structures parallel those observed for (uncharacterised) PTO and STO monolayers (Table 1).

**Table 1.** Bandgaps calculated by PBEsol for monolayers containing 2 or 3 atomic planes.

| Material | PBEsol (eV) |
| --- | --- |
| PbO-TiO$_2$ (PTO) | 2.06 |
| PbO-TiO$_2$-PbO | 1.43 |
| TiO$_2$-PbO-TiO$_2$ | 1.23 |
| SrO-TiO$_2$ (STO) | 1.78 |
| SrO-TiO$_2$-SrO | 1.97 |
| TiO$_2$-SrO-TiO$_2$ | 1.15 |

a: Observed monolayers:[9] PTO 3.5-4.1 eV . STO 3.2-3.5 eV.

Bader atomic charges for all structures considered herein based on PTO and STO monolayers are provided in ESI, and in general these reflect π back bonding contributions of 0.4 electrons to Sr(II), 0.7 to Pb(II), and 2.1 to Ti(IV). These electrons would then become accessible to chemical entities that approach uncoordinated metal sites and hence possibly lead to chemical susceptibilities. This effect could explain the low bandgaps for structures with Ti on both sides, as well as the corresponding stability predicted when Sr is on both sides.

### 3.2 Chemical reactions driving ferroelectricity in monolayers and bilayers made from PTO and STO.

Figure 2 shows optimised structures for both STO and PTO monolayers. For each structure, two views are provided: on top the elevation view looking at the monolayer from the side, and below that the plan view looking at the monolayer from above. In each case, four copies of the optimised $(\sqrt{2} \times \sqrt{2})R45°$ 10-atom unit cell are shown.

Only one stable polymorph of STO is identified, with $p4m$ plane-group symmetry, corresponding to an "octahedral" atomic arrangement; this structure is therefore named **OCT**. The $p4m$ plane group supports a fourfold rotation axis and multiple reflexion planes and hence does not support in-plane polarisation. For PTO, a similar structure is optimised, but it is found to be unstable, with selective Pb-O bond breakage (see Fig. 2) leading to significant energy lowering, resulting in a structure of $p4g$ symmetry. Despite the bond breakage, this structure still supports fourfold rotational symmetry and hence no net in-plane polarisation is induced. Further, Fig. 2 also indicates that this structure is also predicted to be unstable, with small bond-conserving distortions leading to a small energy lowering. The identified polymorph is named **P** and has $pm$ symmetry. As $pm$ symmetry supports only one symmetry plane, in-plane polarisation can be generated in one of the two crystallographic directions.

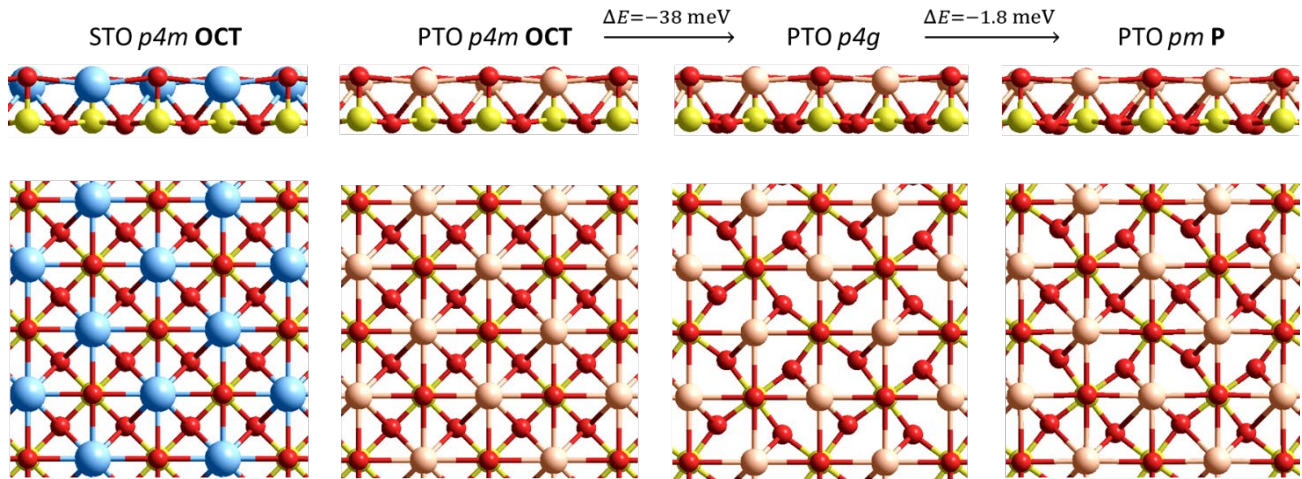

**Figure 2.** Side and top views of four replications of PBESol optimized $(\sqrt{2} \times \sqrt{2})R45°$ structures of STO and PTO monolayers, of descending symmetry: **OCT** structures with *p4m* (square with 4-fold rotation axes and many reflection planes), *p4g* (square with 4-fold rotation axis but less reflection planes) and **P** structures with *pm* symmetry (rectangular, one glide-reflection plane). Reaction electronic energies $\Delta E$ are listed in meV per cell (see Table 2 for the corresponding free energies); Pb- orange, Sr- light blue, Ti- yellow, O- red.

Of note, the optimised (and observed) *x,y* lattice vectors of STO and PTO are similar, 5.433 Å (5.523 Å)[88] and 5.481 Å (5.510 Å),[89] respectively. In both cases, the Ti-O bonding controls the lattice, resulting in nearly commensurate lattices that are suitable for bilayer and heterostructure synthesis. Similar lattice sizes are therefore expected in bilayers, and indeed for STO/PTO, the optimised lattice vectors being *x* = 5.443 Å and *y*= 5.435 Å, compared to an observed[41] value of 5.56 Å.

Next, Fig. 3 provides an overview of the chemical binding changes that can drive in-plane ferroelectricity in bilayers made from STO and PTO. Four types of bilayers can be produced, named "STO/STO", "STO/PTO", "PTO/PTO", and "PTO/STO" to depict the composition of the top layer / bottom layer, as shown in the figure. The figure presents four representations of each structure: a side view, the top view of both layers, and the top view of each individual layer. Visually, the presented structures highlight chemical bonding changes that occur *between* the layers, as well as those that occur *within* the *top* layer.

For STO/PTO, the high-symmetry **OCT** structure forms the stable polymorph. Hence placing an STO monolayer above a PTO monolayer stabilises the intrinsic distortion predicted for PTO. In contrast, the STO/STO bilayer manifest interlayer interactions that drive a small distortion to make $p4g$ symmetry, without inducing in-plane polarisation. Figure 3 shows that this distortion is driven by the formation of new inter-layer Sr-O chemical bonds. For both PTO/PTO and PTO/STO, Fig. 3 shows both the initial **OCT** structures and the structures of the most stable polymorphs identified. For PTO/STO, the most stable polymorph corresponds to the **P** structure of PTO and has *pm* symmetry, whereas for PTO/PTO the most stable polymorph, named **Q**, has rhombic *cm* symmetry with one reflection plane oriented perpendicular to either the *b* or *c* axes.

For STO/PTO and PTO/PTO, MD simulations were performed starting from the lowest-energy structures in Fig. 3, with the energy histories and final structures reported in ESI Figures S1 and S2, respectively. They indicate that the predicted lowest-energy polymorphs are stable on the ps timescale.

The most significant energy lowering reported in Figs. 2-3 is that for PTO/STO. To understand the chemical reactions driving energy lowering, some identified intermediate structures between the initial **OCT** structure, which is a fifth-order saddle point on the potential-energy surface, and various lower-energy structures are shown in Fig. 4. In summary, for $(\sqrt{2} \times \sqrt{2})R45°$ supercells, PBESol predicts low-energy polymorphs **P**, similar but slightly higher-energy structures **Q**, second-order saddle structures named $2S^\theta$ and $2S^\Phi$, and a fourth-order saddle structure $4S^\Phi$. These structures were obtained by distorting **OCT** in some curvilinear phonon modes, described by the historical labels $\theta$ and $\phi$, the significance of which is discussed later in Section 3.5. For non-zero $\theta$, two other significant curvilinear variables $\phi_1'$ and $\phi_2'$ are also identified, the significance of which is discussed later in Section 3.6. These four variables depict aspects of the symmetry lowering (Fig. 4a) from **OCT** that lead eventually to structures supporting in-plane electric polarisation.

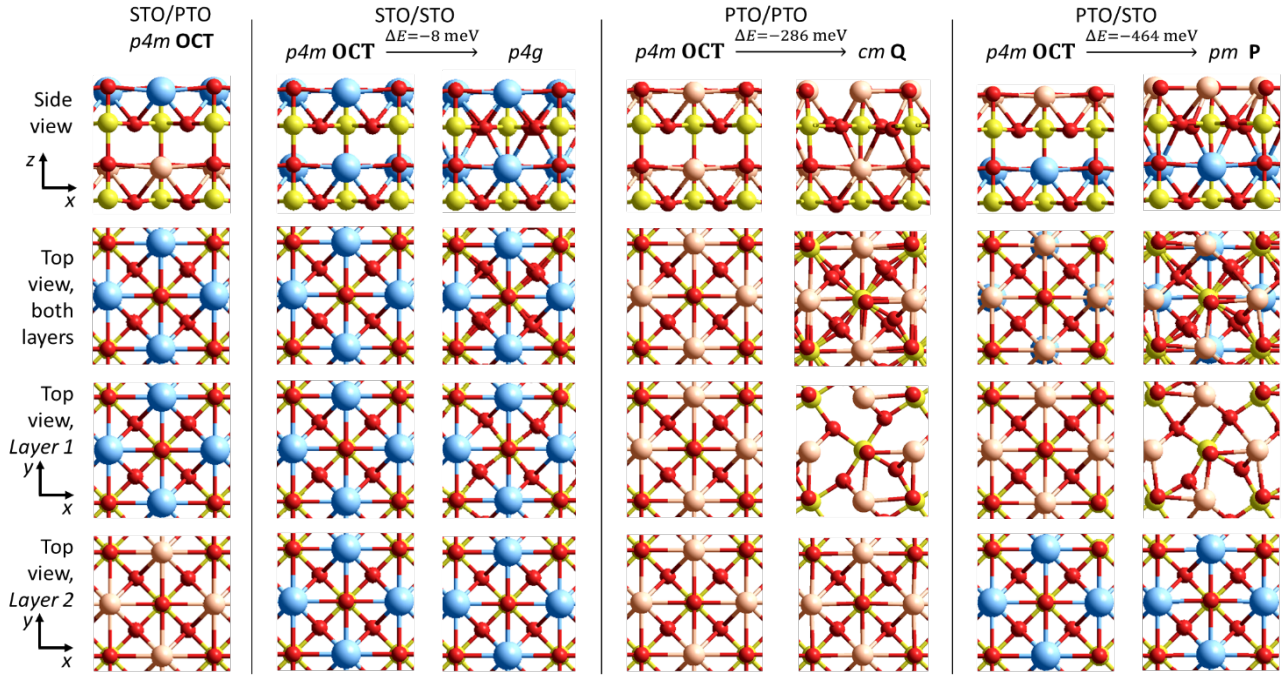

**Figure 3**. Various views of PBESol optimized OCT structures and lowest-energy polymorphs obtained from $(\sqrt{2} \times \sqrt{2})R45°$ representations of STO/PTO, STO/STO, PTO/PTO, and PTO/STO bilayers. Reaction electronic energies $\Delta E$ are listed in meV per cell (see Table 2 for free energies); Pb- orange, Sr- light blue, Ti- yellow, O- red.

All structures except **OCT** can take on equivalent symmetry-related forms depending on the direction in which the phonon distortions occur, with e.g., the most stable polymorphs, named $\mathbf{P}_i$ and $\mathbf{P}_i$, for $i = 1$ to 4, depict eight equivalent structural forms that are related by 90° rotations and/or internal cell translations by (0.5,0.5,0), see Fig. 4b. Similarly, there are two unique but equivalent forms for $\mathbf{4S^\phi}$, four for $\mathbf{2S^\theta}$, eight for $\mathbf{2S^\phi}$, and eight for **Q**.

Compared to **OCT**, the eight structures **P** shown in Fig. 4b manifest the breaking of different instances of two types of intra-layer bonds: Pb-O bonds to one of the four oxygen atoms that lie in the TiO$_2$ plane, and Pb-O bonds to the oxygen that sits above the TiO$_2$ plane. Concerning the chemical representations depicted in Fig. 4, emphasis is placed on the bonds that break within the top (PTO) layer of the bilayer, and so only its structure and internal bonding are shown.

Figure 4c considers the reactions that lead to the breaking of Pb-O bonds to oxygens that lie within the TiO$_2$ plane. Distortion in $\phi$ from **OCT** to $\mathbf{4S^\phi}$ captures the essential features of these chemical reactions, followed by a distortion dominated by $\theta$ to $\mathbf{2S^\phi}$ that breaks some of the alternative Pb-O bonds. Then a third distortion dominated by $\phi_1'$ and $\phi_2'$ breaks more bonds to form the low-energy structures **P** and **Q**. The most important step in terms of energetic changes is the first step, but this does not induce in-plane polarisation as the symmetry of $\mathbf{4S^\phi}$ is $p4g$, supporting fourfold rotational symmetry.

Alternatively, Fig. 4d considers first the distortion dominated by $\theta$ that simultaneously breaks all of the Pb-O bonds to O above the TiO$_2$ plane, producing $\mathbf{2S^\theta}$ structures of symmetry $cmm$. These structures support two orthogonal reflection planes and hence do not manifest in-plane polarisation. Subsequent distortion in $\theta$, which has a larger energetic effect, then, in conjunction with $\phi_1'$ and $\phi_2'$, breaks the in-plane Pb-O bonds to produce the low-energy structures **P** and **Q**.

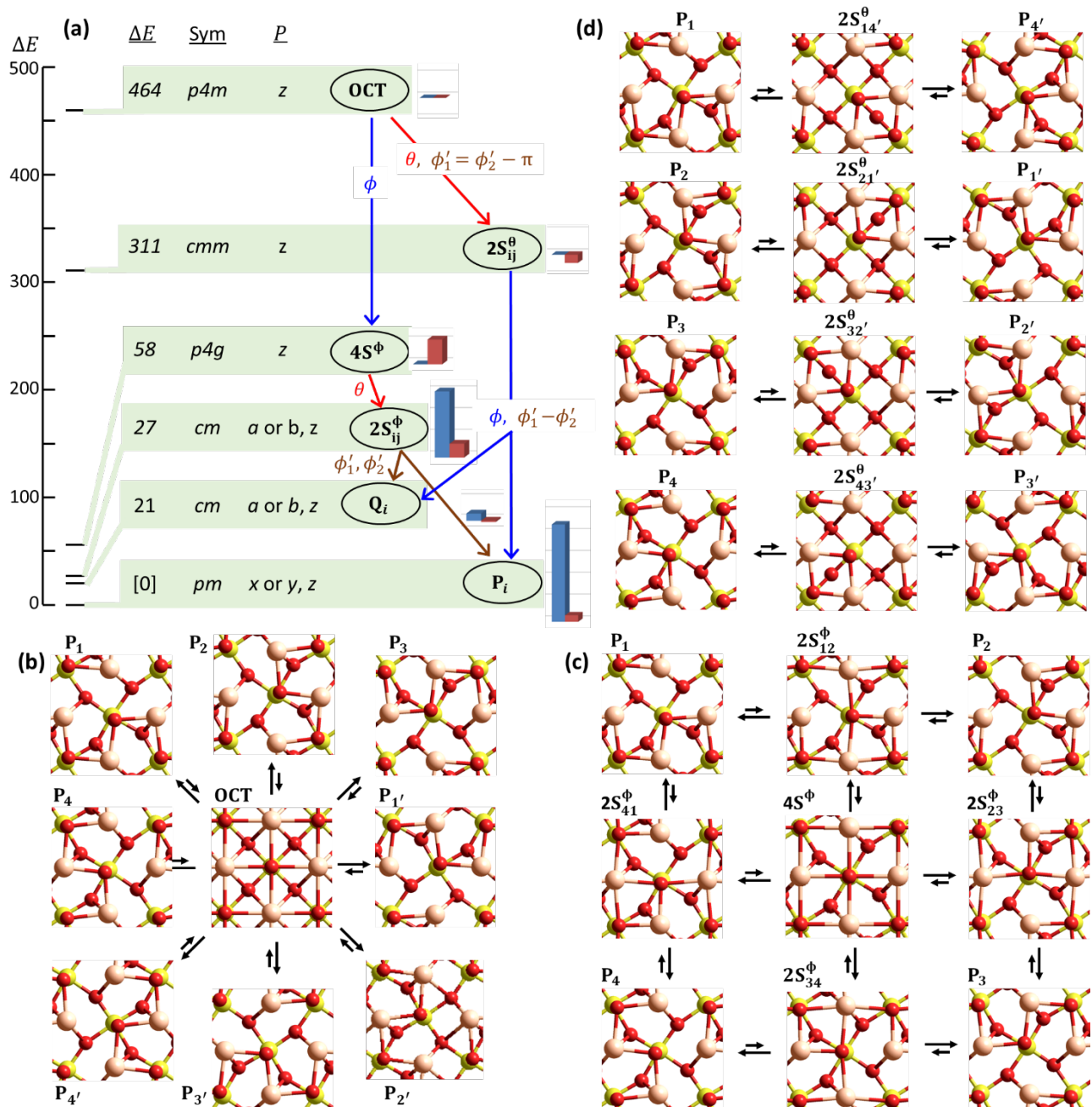

**Figure 4**. Significant PBESol structures pertaining to PTO/STO polymorphism, showing *Layer 1* from optimized $(\sqrt{2} \times \sqrt{2})R45°$ supercells (Pb- orange, Ti- yellow, O- red). The structure labels are: $P_i, P_{i'}$ – most electronically stable polymorphs, $Q_i$ – higher energy polymorphs, $2S_{ij}$- second-order saddle points, $4S$- 4$^{th}$-order saddle structures, and **OCT**- high symmetry (octahedral-like) structure, a 5$^{th}$-order saddle point. Rotationally related phases are numbered $i = 1 - 4$; the structures $1' - 4'$ are translationally equivalent to these. (a) Symmetry, relative electronic energy (in meV per cell, see Table 2 for free energies), allowed polarisation components, and direct-reaction pathway summary; reaction pathways starting with pure distortions in either of the AFD variables $\theta$ (plus also associated polar angles $\phi'_1$ and $\phi'_2$) and $\phi$ are indicated; the bar charts show the relative polarisations to **OCT** (left in blue- in-plane magnitude, right in brown- vertical. (b) Relationship of the various **P** polymorphs to the **OCT** structure. (c) Some low-energy direct polymorphism pathways (locally expanded in Fig. 9c). (d) Some high-energy direct polymorphism pathways (locally expanded in Fig. 9b).

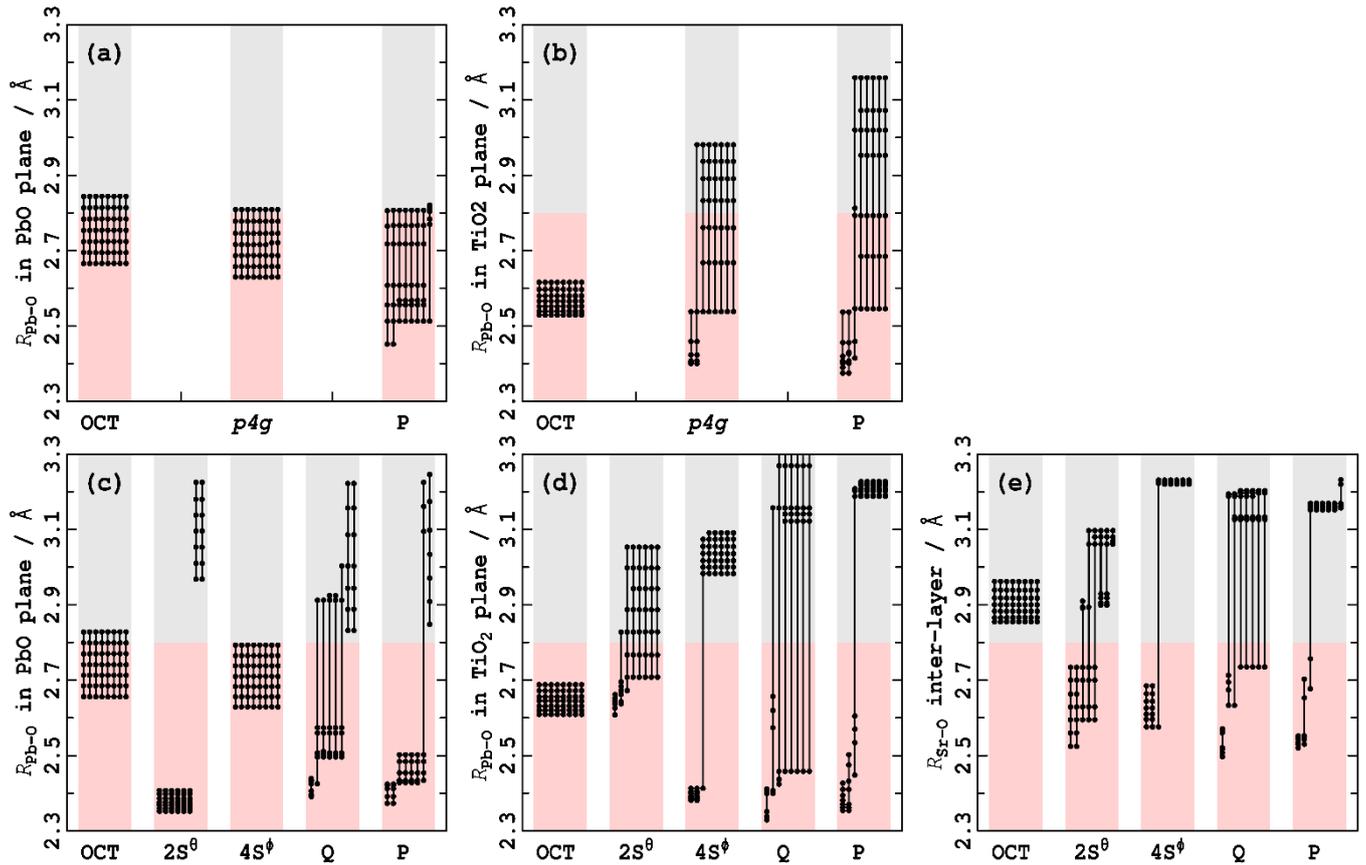

**Figure 5.** Variation of PBESol PTO monolayer (a,b) and PTO/STO bilayer (c,d,e) bond lengths with strain in $(\sqrt{2}\times\sqrt{2})R45°$ supercells. The marked lines connect dots representing one of the eight possible bond lengths as the biaxial strain varies from -3% (shortest bond lengths) to -2%, -1%, 0, 1%, 2%, and finally 3% (longest bond lengths). Pink-shaded regions and grey-shaded regions depict bond lengths flagged as either made or broken, respectively.

### 3.3 Examining the effects of biaxial strain reveals the chemical bonding changes act as binary switches.

Placing materials under strain can perturb their chemical bonding networks. To examining this effect, the monolayers and bilayers made from PTO and/or STO were placed under biaxial strains ranging from -3% to 3%, values that in principle could be achieved by placing the heterostructures on top of appropriate substrates.

Figure 5 provides a graphical display of the 16 possibly unique Pb-O bond lengths in PTO (Fig. 5a,b) and in PTO/STO (Fig. 5c,d). These results are subdivided into two categories, representing bonds to O atoms located in the PbO plane (Fig. 5a,c) and in the TiO$_2$ plane (Fig. 5b,d). Also included are the eight interlayer Sr-O bond lengths (Fig. 5e). Results are presented for the three identified structures of different symmetry for PTO and the five such structures for PTO/STO. The eight bond lengths per frame are connected by lines showing variations of interatomic distance at biaxial strains of -3%, -2%, -1%, 0, 1%, 2%, and 3%. Coloured shading indicates the regions used in the definition of Pb-O and Sr-O bonding.

High-symmetry structures such as **OCT** demand that all bond lengths respond uniformly as a function as biaxial strain, scaled by the proportion of the bond vector that is oriented in-plane and hence responsive to the strain. As a result, bond lengths evolve smoothly between those that could be interpreted as being "made" and those that could be interpreted as being "broken".

For lower-symmetry structures, this is not always the case, however, especially for PTO/STO. Instead, as the biaxial strain increases, many bonds make and break. Note that bond lengths are found that could be considered as being intermediary between "made" and "broken", indicating the robustness of the bonding definition used.

At certain values of biaxial strain, Pb-O and Sr-O bonds change abruptly, making the presence/absence of bonds a binary switching mechanism that operates to control the structures of the monolayers and bilayers. In-plane ferroelectricity results when enough of these binary switches toggle to reduce the symmetry below $p4m$, $p4g$, and $cmm$ to either $cm$ or $pm$ (see e.g., Fig. 4). Chemical bonding effects therefore control ferroelectricity in this bilayer. A similar effect is seen for PTO with some bond lengths changing by amounts much larger than would be generated by a uniform response to the applied strain, but the clear made/broken categorisation found for PTO/STO is not as pronounced.

### 3.4 Quantifying the energetic and polarisation changes.

Table 2 lists the PBESol calculated energy (both the electronic energies $\Delta E$ portrayed in the figures and free-energy changes $\Delta G$), as well as the in-plane polarisation-magnitude ($|\Delta P_{xy}|$) changes for all optimised structures with respect to that of the appropriate **OCT** structure. The free-energy changes are provided at both 0 K and 298 K, based upon obtained vibrational analyses for each structure. Significant roles for the influence of the phonons on structure stability are predicted.

**Table 2.** Structure energy analysis. Electronic energy differences $\Delta E_\text{PBESol}$ and associated free-energy differences $\Delta G_\text{PBESol}$ at 0 K and 298 K, along with energy differences obtained using CAM-B3LYP and CAM-B3LYP/D3(BJ), and hence best-estimate free-energy changes, as well as PBESol in-plane polarisations magnitudes $|\Delta P_{xy}|$, for polymorphs of PTO and STO monolayers and bilayers with respect to **OCT** structures, obtained for $(\sqrt{2} \times \sqrt{2})R45°$ supercells.[a]

| Material | Str. | $\Delta E_\text{PBESol}$ (meV) | $\Delta E_\text{CAM}$ (meV) | $\Delta E_\text{CAM/D3}$ (meV) | $\Delta G^0_\text{PBESol}$ (meV) | $\Delta G^{298}_\text{PBESol}$ (meV) | $\Delta G^0_\text{est}$ (meV) | $\Delta G^{298}_\text{est}$ (meV) | $|\Delta P_{xy}|$ ($\mu C/cm^2$) |
|---|---|---|---|---|---|---|---|---|---|
| PTO | $p4g$ | -38 | <u>-39</u> | -70 | <u>-38</u> | 29 | -70 | -3 | 0 |
|  | P | <u>-39</u> | -33 | <u>-76</u> | -33 | <u>-58</u> | -70 | <u>-95</u> | 4.2 |
| STO/STO | $p4g$ | -8 | -9 | -23 | -21 | -22 | -36 | -37 | 0 |
| PTO/PTO | Q | <u>-286</u> | <u>-343</u> | -267 | <u>-255</u> | <u>-317</u> | -236 | -298 | 2.4 |
|  | P | -233 | -212 | <u>-332</u> | -202 | -264 | <u>-301</u> | <u>-363</u> | 13.5 |
| PTO/STO | $2S^\theta$ | -153 | -178 | -192 | -133 | -203 | -172 | -242 | 0 |
|  | $4S^\phi$ | -406 | -419 | -482 | -396 | -381 | -472 | -457 | 0 |
|  | $2S^\phi$ | -437 | -460 | -508 | -410 | -439 | -481 | -510 | 6.1 |
|  | Q | -443 | -461 | -511 | <u>-428</u> | <u>-580</u> | -496 | <u>-648</u> | 0.7 |
|  | P | <u>-464</u> | -504 | <u>-539</u> | -422 | -503 | <u>-497</u> | -578 | 8.9 |

[a] The result for the lowest-energy structure predicted for materials with more than one possibility is underlined.

For PTO/STO, the vibrational zero-point energy correction changes the PBESsol 0 K prediction from structure **P** to **Q**, with thermal effects amplifying this change. For PTO, the zero-point energy correction results in a change of the prediction from the polarised structure **P** to the unpolarised one ($p4g$), but for it the thermal corrections lead to a restoration of the polarisation. For PTO/PTO, both corrections conserve the prediction of the polarised structure **Q**. The reliability of all of these predictions will depend significantly on the quality of the computational method used, however, including the choice of density functional and the treatment of the vibration Brillion zone.

As a distinctly different approach to computation of the DFT energetics, single-point energy differences, at the calculated PBESol structures, are evaluated using the asymptotically-corrected and dispersion-corrected CAM-B3LYP/D3(BJ) method and included in Table 2. Best-estimate free energies $\Delta G^0_{est}$ and $\Delta G^{298}_{est}$, that combine all calculated data, are also included therein. These best estimates are qualitatively similar to the pure PBESol results for PTO and PTO/STO, but predict a different lowest-energy structure for PTO/PTO.

As a function of applied biaxial strain, the relative PBESol electronic energies $\Delta E$ of the identified structures are shown in Fig. 6a for PTO and Fig. 6c for PTO/STO. The visually distinct curvatures indicate that the bilayer is predicted to have about twice the force constant compared to that of the monolayer, as expected. Typically, the more bonds identified as "broken" or "made" in a structure (Fig. 5), the easier it is for the structure to respond to imposed strain. For PTO/STO, the force constants are similar for each polymorph, with the exception of $2S^\phi$ for which they are increased. The application of bilateral strain is precited to favour **P** compared to **Q**, but this effect is less than the opposing zero-point motion effects listed in Table 2.

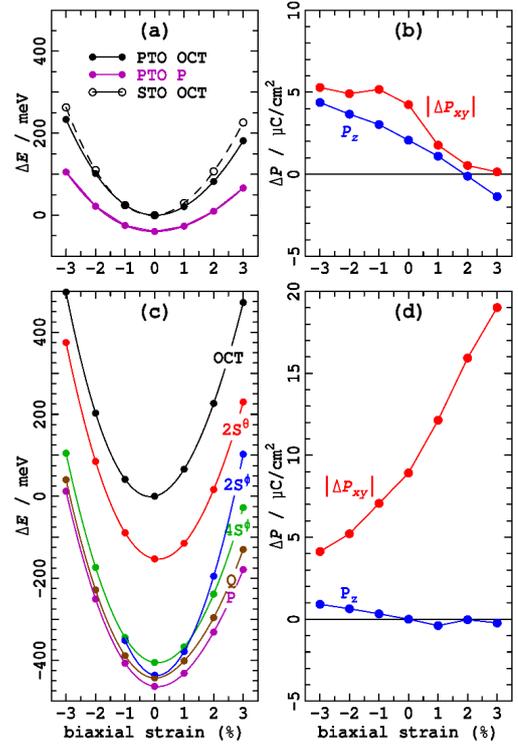

**Figure 6.** The effects of strain on PTO/STO bilayers calculated using PBESol. (a,c) Electronic energy change $\Delta E$ from that of the unstrained **OCT** structures, in meV per $(\sqrt{2} \times \sqrt{2})R45°$ cell, for PTO (a) and PTO/STO (c). (b,d) Polarisation-change $\Delta P$ of polymorphs **P** from that of the unstrained **OCT** structures for (b) PTO and (d) PTO/STO (1 $\mu C/cm^2 \equiv 0.146$ eÅ/cell).

The magnitudes of the in-plane polarisations induced in structures **P** (Table 2) increases from 4.2 μC/cm$^2$ for PTO to 13.5 μC/cm$^2$ for PTO/PTO and 8.9 μC/cm$^2$ for PTO/STO. These values are similar to those reported for 2D SnTe[15] of 8-12 μC/cm$^2$ and bulk BaTiO$_3$[90] of 10-20 μC/cm$^2$.

In Fig. 6b and Fig. 6d, the calculated in-plane and out-of-plane polarisation magnitudes are shown for PTO and PTO/STO, respectively. As the operation of chemical reactions acting as binary switches determines the effect of strain (Fig. 5), the shown curves are not always smooth. Nevertheless, the out-of-plane polarisation generally decreases with increasing strain, as is expected as strain effectively reduced the width of each layer. For PTO, the in-plane polarisation also mostly decreases with increasing strain, but for PTO/STO, increasing strain induces a large increase in the polarisation, achieving the significant polarisation of 19 μC/cm$^2$ at 3% strain. The large effect on the in-plane polarisation predicted as a result of the binding of a STO monolayer below PTO indicates that PTO is sensitive to its chemical surroundings, and indeed Fig. 5e shows that the inter-layer Sr-O bond formation is very sensitive to strain. This effect may be exploitable for the tailoring of device properties involving PTO or for its application in sensing and related technologies.

### 3.5 Chemical bonding changes control AFD and then FE distortions.

Traditionally, induced ferroelectric polarisation in perovskite materials has been associated with lattice and structural distortions[1] of type FE that move atoms asymmetrically, focusing also on the effects of AFD distortions that instead act symmetrically.[7, 16, 17, 25-27, 91] The in-plane polarisation $\Delta P_{xy}$ is often discussed in terms of in-plane displacement vectors $\Delta T_{xy}$ for each atom type, here $\Sigma T_{xy}^{Pb}$, $\Sigma T_{xy}^{Sr}$, $\Sigma T_{xy}^{Ti}$, and $\Sigma T_{xy}^{O}$. In ESI Part S2, a (non-unique) decomposition of the induced polarisation is presented in terms of polarisation induced by atomic translations (proper ferroelectricity) and polarisation induced by electronic redistributions (improper ferroelectricity). This shows that proper ferroelectricity dominates the polarisation and hence establishes the usefulness of such simple descriptions.

The AFD distortions are usually[7, 13] categorised in terms of two effects denoted by changes in the polar angles $\theta$ and $\phi$ describing the spatial orientations of TiO$_6$ octahedra. Definitions of these quantities pertinent to the consideration of monolayers and bilayers are presented in Fig. 7. There, $\theta$ describes the tilting of the top Ti-O bond with respect to the vector normal to the 2D plane, whereas $\phi$ describes in-plane rotation of the central TiO$_2$ sub-layer.

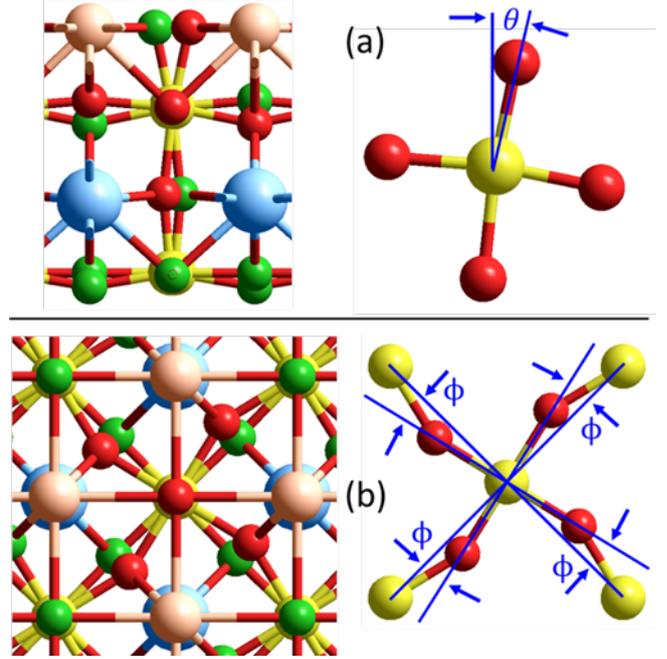

**Figure 7**. Variables $\theta$ and $\phi$ depicting AFD distortions (exampled for PTO/STO; red- O atoms the TiO$_6$ octahedron of interest, green- other O; Pb-orange, Sr- blue, Ti- yellow). (a) Side view of $2S^\theta$, showing tilting of an octahedral unit about a horizontal axis; as the lower oxygen is absent in bottom layers, the angle $\theta$ is taken simply as the orientation of the upper Ti-O bond to the vertical. (b) Top view of $4S^\phi$, highlighting (perhaps inequivalent) in-plane twisting angles $\phi$.

From Fig. 4, it is clear that the distortion from **OCT** that drives the greatest reduction in energy for PTO/STO is that associated with the AFD motion in $\phi$, which represents the chemical reactions that convert **OCT** to $4S^\phi$ and dominates the conversion of $2S^\theta$ to either **P** or **Q**. Motion in $\phi$, e.g., the conversion of OCT to $2S^\theta$, is the next most significant effect.

Acting alone, neither of these effects induces dipole polarisation as both are AFD in nature. Nevertheless, when acting *together*, they lower the symmetry sufficient to allow for ferroelectric polarisation. Acting together permits FE distortions to occur, distortions that have minimal effect on energetics but profound impact concerning polarisation. The net effect of the chemical bonding changes is therefore to generate FE displacements $\Sigma T_{xy}$. Utilising data for PTO/STO structures **P** (under strain), **Q**, and $2S^\phi$, the unsigned magnitudes of these displacements are correlated with the total in-plane polarisation magnitude $|\Delta P_{xy}|$ in Fig. 8. The dominant contribution (~ 50%) to polarisation in PTO/STO arises from the net effect of the 12 oxygen atoms per unit cell, but per atom the major contribution comes from the two Pb atoms per cell. In addition, the four Ti atoms make a small contribution but the contribution from the two Sr atoms is negligible. The AFD distortions in $\theta$ and $\phi$ combine to strengthen and weaken Pb-O

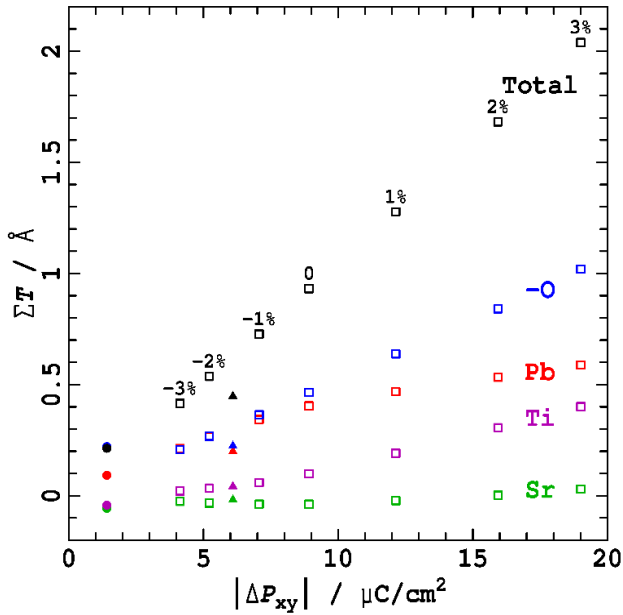

**Figure 8.** Shown for PTO/STO structures is the calculated correlation of the change of the in-plane polarisation magnitude from OCT, $|\Delta P_{xy}|$, with the symmetric in-plane atom translations $\Sigma T^k_{xy}$, for: open squares- **P** at various levels of biaxial strain, filled triangles- $2S^\phi$, and filled circles- **Q**. Shown properties include the total polarisation in black and its partitioning into contributions from different atom types: blue- oxygen (negative sign, 12 atoms per cell), red- Pb (two atoms per cell), purple- Ti (four atoms per cell), and green- Sr (two atoms per cell).

and Sr-O bonds (Fig. 5), resulting in net atomic displacements that induce proper ferroelectricity.

### 3.6. How bonding changes combine to control polarisation switching.

Previously, Fig. 4 showed distortions in the critical curvilinear variables $\theta$ and $\phi$ that generate series of second-order saddle points of types $2S^\theta$ and $2S^\phi$ that lead on to polymorphs **P** and **Q**. With variations of density functional, zero-point and thermal corrections, Table 2 showed changes in the relative ordering of **P** and **Q**, and hence the expected ground-state polymorph polarisation, but the primary qualitative scenario remains. In principle, **P** and **Q** could represent stable polymorphs or interconnecting transition states; if both are stable polymorphs then indeed there must exist some intermediary transition state that would be critical to polarisation switching. Here, we investigate the nature of polarisation switching of PTO/STO, as perceived by PBESol.

As mentioned before **P** is calculated by PBESol to be the most stable polymorph, with **Q** as a metastable polymorph of energy close to that of the second-order saddle $2S^\phi$. The lowest vibration frequency of **Q** is near zero, however, on the borderline of becoming a transition state. Figure 9a defines the two angles $\phi'_1$ and $\phi'_2$ mentioned in Fig. 4, these being the polar angles associated with the distortion of the two vertical Ti-O bonds in the top PTO plane. They are both associated with the tilting of these bonds by angle $\theta$ (see Fig. 6). Then Fig. 9b shows the potential-energy surface around saddle-point $2S^\theta_{14'}$ in variables $\phi$ and $\phi'_2$, whilst Fig. 9c shows that around point $2S^\phi_{12}$ in variables $\phi'_1$ and $\phi'_2$. $2S^\theta_{14'}$ leads to 4 polymorphs $P_1$ and $P_4'$, as indicated in Fig. 4d, and as well $Q_1$ and $Q_{4'}$.

The energy profile along an interpolated reaction coordinated connecting the four polymorphs surrounding $2S^\theta_{14'}$ is shown in Fig. 9d, revealing extremely shallow transition states connecting $P_1$ to $Q_1$ and $P_{4'}$ to $Q_{3'}$, as well as high-energy transition states linking to $Q_1$ to $P_4$, and $P_1$ to $Q_{3'}$. As the energy of the later transition state is only slightly less than that for the (unpolarised) saddle point, reaction around this pathway embodies very large changes in the in-plane polarisation of PTO/STO. This mechanism is not predicted by PBESol to be observable, however, as the energy of the later transition state is too high.

Alternatively, the energy profile along an interpolated reaction coordinated connecting the four polymorphs surrounding $2S^\phi_{12}$ is shown in Fig. 9e and involves only low-energy transition states. The four structures surrounding $2S^\phi_{12}$ are $P_1$ and $P_2$, as indicated in Fig. 4c, and as well $Q_1$ and a previously undescribed structure $R_1$. This new structure appears as a transition state at an energy just below that of the saddle point. The in-plane polarisations of these structures vary smoothly with reaction coordinate, with the pathway from $P_1$ to $P_2$ via $Q_1$ allowing for a 90° change in polarisation at small energy cost. Analogous stepwise conversions to $P_3$ and $P_4$ that move around other instances of $2S^\phi$ (see Fig. 4c) then allow for polarisation to be moved by further 90° increments.

The identified transition-state $R_1$ has a very flat potential-energy surface in the direction of $2S^\phi_{12}$ and so the optimised structure has less distortion in $\phi'_1$ and $\phi'_2$ than expected. In Fig. 9, $R_1$ is shown as it its distortion had the same magnitude as that found for $P_1, P_2, Q_1$, revealing its intrinsic bonding nature. Inside the top Pb-O plane of PTO/STO, the oxygen atoms distort from their **OCT** positions to make bonds to two of the neighbouring Pb atoms. In the structures **P**, these bonds form either horizontal or vertical lines, whereas in the structures **Q** they make isolated squares, and in the structures **R** they make diagonal lines. Independent of the computational methods used, these bonding patterns will dominate the perceived polarisation and its switching mechanism.

## 4. Conclusions

Bulk materials and extensive heterostructures made of STO and PTO layers present considerable applications interest as they display ferroelectricity, as well as possibly type-II band alignment. These features offer photocatalytic and photovoltaic possibilities that could perhaps be controlled using the ferroelectric properties. In terms of basic understanding, the central issues to be resolved concerned how ferroelectricity is generated. By considering only simple structures such as monolayers and bilayers, it is revealed that the drive to optimise intra-layer Pb-O and inter-layer Pb-O or Sr-O bond lengths controls energetics, structure and polarisation (Figs. 4 and 5). Polymorphism arises as structures evolve to maximise the number of strong bonds, at the expense of breaking weaker ones. Ground-state polarisation switching involves reactions that concertedly make and break bonds (Fig. 9).

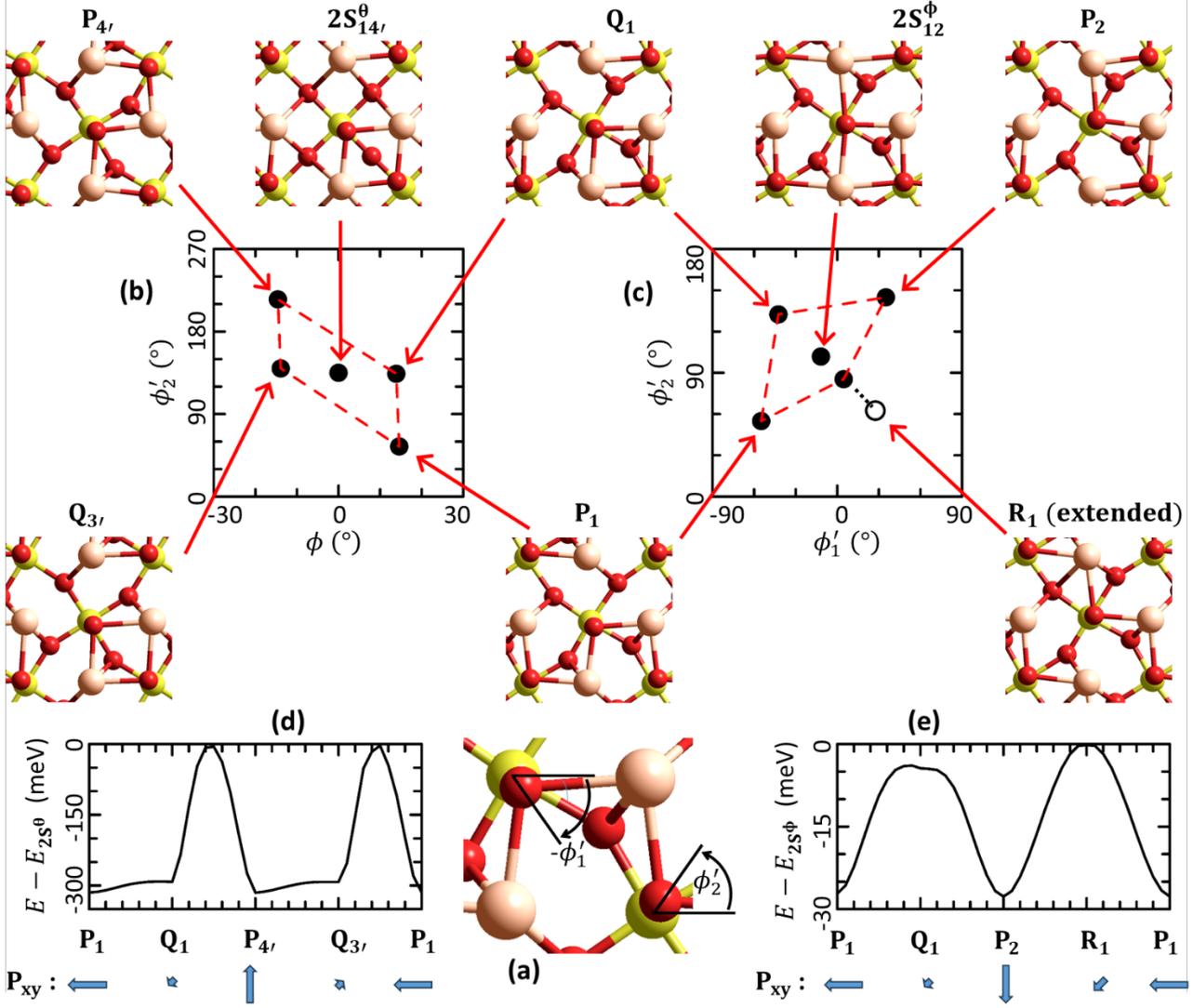

**Figure 9.** The angular distortions that depict revealing the energy profile for polarisation switching. (a) The polar angles $\phi'_1$ and $\phi'_2$ associated with $\theta$ that depict the orientation of the nearly vertical Ti to O bonds. (b) The structures surrounding $2S^{\theta}_{14'}$: $P_1$, $Q_1$, $P_{4'}$ and $Q_{3'}$ as a function of $\phi$ and $\phi'_2$. (c) The structures surrounding $2S^{\phi}_{12}$: $P_1$, $Q_1$, $P_2$, and $R_1$ (depicted at the same extent of distortion as found for $P_1$, $Q_1$, $P_2$) as a function of $\phi'_1$ and $\phi'_2$. (d) Interpolated potential-energy surfaces for traverse around $2S^{\theta}_{14'}$, including the in-plane polarisation vectors for the key structures. (e) Interpolated potential-energy surfaces for traverse around $2S^{\phi}_{12}$, including the in-plane polarisation vectors for the key structures.

The bonding effect drives primarily the AFD distortions of local TiO$_6$ octahedral through distortions in either $\theta$ (tilting) or $\phi$ (in-plane twisting), as depicted in Fig. 8. Enabling either distortion can significantly lower the energy of the system, but either distortion acting alone fails to produce ferroelectricity. When both distortions occur simultaneously, they compete with each other, with a side-effect being the manifestation of FE distortions that induce in-plane polarisation. It is the bonding-driven competition between the two primary AFD effects that generates the FE ones. This competition manifests through the variables $\phi'_1$ and $\phi'_2$ (Fig. 9): these curvilinear variables are generated through distortion in $\theta$, with $\phi$ controlling the values that they can take. Together, these variables describe the concerted bond-breaking and bond-making processes that control polarisation switching.

Even though the lattice vectors of PTO and STO are predicted to be very similar, a significant difference between PTO and STO layers is found to be the optimal bond lengths to oxygen, which are ~ 0.3 Å longer for Sr than for Pb. Hence, for all structures, the overall size appears to be controlled by the Ti-O bonding network, within which the intra-layer Sr-O bonds are optimised but the Pb-O bonds are too short. This leads to STO layers being stable in their high-symmetry **OCT** forms, whereas PTO ones are not and so spontaneously undergo AFD distortions. In addition, in OCT structures both inter-layer Sr-O bonds and Pb-O bonds are too short. Nevertheless, in the STO/PTO bilayer, the binding of the STO onto the Pb atoms inhibits the intrinsic PTO distortion in the Pb-O bonds, whereas in PTO/PTO or PTO/STO bilayers, the addition of a poorly-bonded layer below the PTO enhances the distortion. These basic effects would be expected to operate in all mixed-layer solids and heterostructures made using PTO and STO, providing control over properties and function.

The net result is the heterostructures made from PTO and STO are expected to be controlled by is a fine energy balance between different properties, including subtle bonding rearrangements, inter-layer interactions, strain, zero-point motion and thermal effects. As a result, large changes in in-plane polarisation can result from seemingly small perturbations at minimal energy cost, leading to a useful, highly polarizable system.

## Conflicts of interest

There are no conflicts to declare.

## Acknowledgements

We thank for the National Natural Science Foundation of China (12347115, 12074241, 11929401, 52130204), the Science and Technology Commission of Shanghai Municipality (19010500500, 20501130600, 22XD1400900), High Performance Computing Centre, Shanghai University, National Computational Infrastructure (Australia), the Major Scientific Project of Zhejiang Lab (No. 2021PE0AC02), and the Australian Research Council Centre of Excellence for Quantum Biotechnology (grant No. CE230100021). Dr. Jia appreciated the support of China Postdoctoral Science Foundation (No. 2022M722035). For computing resources we than the Shanghai University ICQMS facility and National Computational Infrastructure Australia NCMAS grant d63 augmented by funding from the University of technology Sydney.

## Notes and references